\documentclass{article}

\usepackage{jheplike}
\usepackage[T1]{fontenc}
\usepackage{amsmath}
\usepackage{xcolor}
\usepackage{caption} 
\usepackage[htt]{hyphenat}

\usepackage[shadow,textwidth=2.7cm]{todonotes}

\allowdisplaybreaks
\title{Feynman parameter integration through differential equations}

\preprint{UUITP-31/22\\
\hspace*{0pt}\hfill{} CERN-TH-2022-111}

\author[1]{Martijn Hidding,}
\author[2]{Johann Usovitsch}

\affiliation[1]{Department of Physics and Astronomy, Uppsala University, SE-75120 Uppsala, Sweden}

\affiliation[2]{Theoretical Physics Department, CERN, 1211 Geneva, Switzerland}

\emailAdd{martijn.hidding@physics.uu.se}
\emailAdd{johann.usovitsch@cern.ch}

\abstract{We present a new method for numerically computing generic multi-loop Feynman integrals. The method relies on an iterative application of Feynman's trick for combining two propagators. Each application of Feynman's trick introduces a simplified Feynman integral topology which depends on a Feynman parameter that should be integrated over. For each integral family, we set up a system of differential equations which we solve in terms of a piecewise collection of generalized series expansions in the Feynman parameter. These generalized series expansions can be efficiently integrated term by term, and segment by segment. This approach leads to a fully algorithmic method for computing Feynman integrals from differential equations, which does not require the manual determination of boundary conditions. Furthermore, the most complicated topology that appears in the method often has less master integrals than the original one. We illustrate the strength of our method with a five-point two-loop integral family.}

\begin{document}

\maketitle

\section{Introduction}

The computation of multi-loop Feynman integrals is a crucial component in generating predictions for particle processes at high-energy colliders such as the LHC. Many modern techniques for computing Feynman integrals rely on the differential equation method \cite{Kotikov:1990kg, Henn:2013pwa}. This method involves setting up a system of differential equations for the master integrals of a Feynman integral family, and solving the differential equations either analytically or numerically. Analytic solutions can often be found in terms of special classes of iterated integrals such as multiple polylogarithms \cite{Goncharov:1998kja,Goncharov:2001iea} or generalizations thereof (see e.g. \cite{Broedel:2017siw, Adams:2018yfj}), but are hard to obtain in general. In addition, there are many Feynman integrals which do not evaluate to well-studied classes of iterated integrals. Therefore, renewed interest has been expressed in solving differential equations numerically without reference to an intermediate function space.

Many works in the literature have studied methods for solving differential equations for Feynman integrals numerically (see e.g. \cite{Pozzorini:2005ff, Liu:2017jxz, Lee:2017qql, Mandal:2018cdj, Czakon:2020vql}). In this paper, we employ the strategy outlined in \cite{Moriello:2019yhu}. In this approach, differential equations are repeatedly solved in terms of generalized power series expansions (which may contain powers of logarithms) along connected line segments in phase-space. The Mathematica package \texttt{DiffExp} \cite{Hidding:2020ytt} provides a public implementation of the method that works on user-provided differential equations. (Another recent public implementation is the \texttt{SeaSyde} \cite{Armadillo:2022ugh} Mathematica package.) The approach yields a precise and efficient way of computing Feynman integrals at any point in phase-space, given an initial set of boundary conditions as input. The computation of boundary conditions can be performed analytically in special limits with the aid of expansion by regions \cite{Beneke:1997zp, Smirnov:1999bza, Jantzen:2011nz}, but generally requires some manual work.

In \cite{Liu:2022chg} the Mathematica package \texttt{AMFlow} was presented which is based on the method of auxiliary mass flow \cite{Liu:2017jxz, Liu:2021wks}. The package uses expansion by regions to automatically determine boundary conditions in a special limit of infinite complex mass, and a customized series solution solver to transport the boundary conditions to physical points in phase-space. The result is an efficient and fully automated package for computing Feynman integrals numerically at high precision. In \cite{Dubovyk:2022frj} we explored a different automated strategy which relies on sector decomposition \cite{Binoth:2000ps,Borowka:2018goh,Smirnov:2021rhf} to compute boundary conditions in the Euclidean region for a special basis of (quasi-)finite master integrals \cite{vonManteuffel:2014qoa}. We then obtained results in physical regions by transporting the boundary conditions with \texttt{DiffExp}. This strategy is fully automated as well, but it is not able to obtain results at a precision as high as \texttt{AMFlow}. 

The main result of the present paper is a new automated approach for computing Feynman integrals numerically, which relies on repeatedly combining two propagators using Feynman's trick:
\begin{align}
    \label{eq:feyntrick}
    \frac{1}{D_i^{\nu_i} D_j^{\nu_j}} = \frac{\Gamma(\nu_i+\nu_j)}{\Gamma(\nu_i)\Gamma(\nu_j)}\int_0^1 dx\, \frac{ x^{\nu_i-1}(1-x)^{\nu_j-1}}{(D_i x+ D_j (1-x))^{\nu_i+\nu_j}}\,,
\end{align}
where $\nu_i$ and $\nu_j$ are assumed to be positive integers. We will show that the integration with respect to the Feynman parameter $x$ can be performed by solving an associated system of differential equations in terms of generalized series expansions, and by integrating the series expansions term by term. This method has the computational advantage that the first application of Feynman's trick typically gives a simplified integral family with less master integrals than the original family \cite{Papadopoulos:2019iam}. This is generally not the case for the auxiliary mass flow method, which introduces (complex) masses on the propagators. The reduction of the number of master integrals in our approach simplifies the integration by parts (IBP) reduction, which can present a bottleneck in complex calculations.

We remark that the use of \eqref{eq:feyntrick} was also considered in \cite{Hidding:2017jkk} and \cite{Papadopoulos:2019iam} in the context of simplifying Feynman integral computations. In \cite{Hidding:2017jkk}, the trick was used to study differential equations of elliptic Feynman integrals by deferring the elliptic type integration to the last integration parameter. In \cite{Papadopoulos:2019iam}, the formula was shown to simplify an integral family when the combined propagators $D_i$ and $D_j$ have the same internal loop momentum, and the method was called the \lq internal reduction method\rq. Furthermore, it was already observed in \cite{Papadopoulos:2019iam} that this procedure reduces the number of master integrals. 

This paper expands on previous ideas in a number of novel ways. Firstly, we discuss how \eqref{eq:feyntrick} can be employed recursively, and how the necessary integrations can be performed by solving differential equations with series expansion methods. Furthermore, we present a formula for resolving poles in dimensional regularization which is needed to regulate the integrand, and we discuss how to apply it to an integrand which is in a generalized series representation. We also discuss some first steps into obtaining $i\delta$-prescriptions for the Feynman parameters, in order to obtain results outside the Euclidean region.

The remainder of the paper is organized as follows. In section \ref{sec:preliminaries} we set up conventions, and we review the Feynman parametrization and Feynman's trick for combining propagators. In section \ref{sec:directintegrationthroughdiffeqns} we describe the heart of our method, which is to perform the integrals in \eqref{eq:fundamentalrecursion} by making use of differential equations and series solution methods. We describe in section \ref{sec:ftapproachregularization} how to regulate the Feynman parameter integrals in our approach when the master integrals are not finite in $\varepsilon$, and in section \ref{sec:threshold} we discuss first steps into crossing threshold singularities in our approach, in order to obtain results directly in physical regions. Lastly, we discuss in section \ref{sec:example} two non-trivial pedagogical two-loop examples which serve as a proof of concept for our method. Finally, we summarize our methods and results in section \ref{sec:conclusion}.

\section{Preliminaries}
\label{sec:preliminaries}
Here we review a few general definitions of Feynman integrals. For a thorough review, see e.g. \cite{Weinzierl:2022eaz}. The main object of interest is a family of Feynman integrals given by
\begin{align}
\label{eq:integralfamilydef}
I_{\nu_{1}, \ldots, \nu_{n}}\left(\{s_i, m_i\}, d\right)=\int \left(\prod_{j=1}^{l} \frac{d^{d} k_{j}}{i \pi^{\frac{d}{2}}}\right)\,\prod_{j=1}^{n} \frac{1}{D_j^{\nu_{j}}}\,.
\end{align}
Here $\{s_i, m_i\}$ schematically denotes the set of external scales and internal masses on which the integral family depends. We work in dimensional regularization $d = d_{\mathrm{int}} - 2\varepsilon$, where $d_{\mathrm{int}}$ is a positive (even) integer. In the remainder of this paper, we will typically leave the dependence on the external scales, masses and dimension implicit. The propagator exponents $\nu_i$ are assumed to be integers, and we will denote their sum by $\nu = \nu_1 + \ldots + \nu_n$. Typically some of the exponents are taken to be negative, which gives numerator factors. We use the convention $D_j = -q_{j}^{2}+m_{j}^{2}-i\delta$, where $q_j$ denotes the momentum of the $j$-th propagator. The momentum $q_j$ can be written as a linear combination of $l$ loop momenta $k_i$ and $E$ external momenta $p_j$, such that
\begin{align}
    q_j = \sum_{n=1}^l a_{jn} k_n + \sum_{n=1}^E b_{jn} p_n\,,
\end{align}
for some integers $a_{jn}$ and $b_{jn}$. The Feynman $i\delta$-prescription tells us how to deform around poles in the integration region, and can be viewed as adding an infinitesimally small complex mass in the lower half-plane to each propagator.

It is well known that IBP relations exist between Feynman integrals whose propagator exponents are related by integer shifts \cite{Chetyrkin:1981qh,Laporta:2000dsw}. Furthermore, Feynman integrals in different dimensions may be related through dimensional raising and lowering relations \cite{Tarasov:1996br,Lee:2009dh}.  In order to perform IBP reductions in closed-form, the propagators $D_j$ should form a basis for the span of dot products of the form $k_i \cdot k_j$, and $k_i \cdot p_j$. A minimal set of independent integrals may be found in terms of which all other integrals can be expressed, which is called a basis of master integrals. Finding the basis of master integrals can be done through IBP-reduction programs such as LiteRed \cite{Lee:2013mka}, Reduze \cite{vonManteuffel:2012np}, \texttt{FIRE} \cite{Smirnov:2019qkx} and \texttt{KIRA} \cite{Maierhoefer:2017hyi}. The computational complexity of the IBP reductions is typically closely connected to the number of master integrals that the integral family has.

\subsection{Feynman parametrization}
It is common to write Feynman integrals in the so-called Feynman parametrization. For brevity, we will assume in the following that all propagator exponents $\nu_1,\ldots,\nu_n$ are positive, and that we do not necessarily have a basis of propagators and numerators suitable for IBP reduction. For the case with numerators, we refer the reader to \cite{Lee:2013mka}. First, we consider the following formula, which was first observed by Feynman, and which we will refer to as Feynman's trick:
\begin{align}
    \label{eq:feyntrickgeneral}
    \frac{1}{D_1^{\nu_1} \ldots D_n^{\nu_n}} = \frac{\Gamma(\nu)}{\Gamma(\nu_1)\ldots\Gamma(\nu_n)}\int_0^1 d^n \vec{x}\, \frac{x_1^{\nu_1-1} \ldots x_n^{\nu_n-1}\delta\left(1-\sum_{j=1}^{n} x_{j}\right)}{(x_1 D_1 + \ldots + x_n D_n)^{\nu}}\,,
\end{align}
where $\nu = \nu_1 + \ldots + \nu_n$. Upon plugging the formula into \eqref{eq:integralfamilydef}, we may integrate out the loop momenta, which gives the Feynman parametrization:
\begin{align}
\label{eq:feynmanparametrization}
I_{\nu_{1}, \ldots, \nu_{n}}=\frac{\Gamma(\nu-l d / 2)}{\prod_{j=1}^{n} \Gamma\left(\nu_{j}\right)} \int \left(\prod_{j=1}^{n} d x_{j}\, x_{j}^{\nu_{j}-1}\right) \frac{\mathcal{U}^{\nu-(l+1) d / 2}}{\mathcal{F}^{\nu-l d / 2}}\delta\left(1-\sum_{j=1}^{n} x_{j}\right)\,.
\end{align}
The graph polynomial $\mathcal{U}$ is of degree $l$ in the Feynman parameters $x_i$, and $\mathcal{F}$ is of degree $l+1$. These are also called the first and second Symanzik polynomials. The Symanzik polynomials can be constructed from the graph $G$ of the Feynman integral. It holds that
\begin{align}
\mathcal{U}&=\sum_{T \in T(G)} \prod_{e_{i} \notin T} x_{i} \,,\quad\quad \mathcal{F}_0=\sum_{\left(T_{1}, T_{2}\right) \in T_{2}(G)}\left(\prod_{e_{i} \notin\left(T_{1} \cup T_{2}\right)} x_{i}\right) s_{\left(T_{1}, T_{2}\right)}, \quad \nonumber \\
\mathcal{F}&=-\mathcal{F}_0+\mathcal{U}\left(\sum x_{i} m_{i}^{2}\right)-i\delta\,.
\label{eq:symanzikdef}
\end{align}
The sum goes over the spanning trees $T(G)$ of $G$, and the spanning 2-forests $T_2(G)$ of $G$. We have added a $-i\delta$ to $\mathcal{F}$, which takes care of the Feynman prescription.

More generally, we may wish to consider Feynman integrals with generalized propagators, which are quadratic functions in the internal and external momenta. In this case a graph representation may not exist for the Feynman integral, but the Symanzik polynomials can still be defined through determinants. We refer the reader to \cite{Bogner:2010kv, Weinzierl:2022eaz} for explicit formulas.

\subsection{Iterating Feynman's trick}
\label{sec:iterateFeynmanTrick}
Let us derive an alternative form of the Feynman parametrization by repeatedly applying \eqref{eq:feyntrickgeneral} to combine two propagators. For example, we may choose to combine every time the leftmost two propagators, which we will do next. More generally, we may choose to combine propagators in some other sequence, which we will do for our examples. Let us consider the following set of generalized propagators
\begin{align}
    D_{12} &= x_1 D_1 + (1-x_1) D_2\,,\nonumber\\
    D_{123} &= x_2 D_{12} + (1-x_2) D_3\,,\nonumber\\
    \ldots \nonumber\\
    D_{1\ldots n} &= x_{n-1} D_{1\ldots (n-1)} + (1-x_{n-1})D_n\,.
\end{align}
Next, we define a set of integral families $I_{\vec{\nu}}^{(\kappa)}$, such that for $\kappa = 0$ we have the original family, and for each successive step $\kappa$ we combine two propagators. Like before, we will assume in the following that all propagator exponents are positive, and we leave out the numerator factors which are needed to obtain a basis for IBP reductions. We define the integral families by
\begin{align}
I_{\nu_{1}, \ldots, \nu_{n-\kappa}}^{(\kappa)}&=\int\left(\prod_{j=1}^{l} \frac{d^{d} k_{j}}{i \pi^{\frac{d}{2}}}\right) D_{1 \ldots (\kappa+1)}^{-\nu_{1}} \prod_{j=\kappa+2}^{n} D_{j}^{-\nu_{j-\kappa}} \quad \text { for } 0\leq \kappa<n-2, \nonumber\\
I_{\nu}^{(n-1)}&=\int\left(\prod_{j=1}^{l} \frac{d^{d} k_{j}}{i \pi^{\frac{d}{2}}}\right) D_{1 \ldots n}^{-\nu}\,.
\end{align}
We have chosen our conventions such that the integrals $I_{\vec{\nu}}^{(\kappa)}$ depend on the (unintegrated) Feynman parameters $x_1,\ldots,x_{\kappa}$. We remark that an integral  $I_{\vec{\nu}}^{(\kappa)}$ can be represented as a Feynman diagram if we combine propagators with the same internal momentum \cite{Papadopoulos:2019iam}. For example, suppose that
\begin{align}
    D_1 = -(k+p)^2 + m_1^2\,,\quad\quad D_2 = -(k+q)^2 + m_2^2\,.
\end{align}
Then we have 
\begin{align}
    \label{eq:completingthesquare}
    x D_1 + (1-x) D_2 = -(k+P)^2 + M^2\,,
\end{align}
where $P = x p+ (1-x)q$ and $M^2 = x m_1^2 + (1-x)m_2^2 - x(1-x) (p-q)^{2}$. Note that $I_{\nu}^{(n-1)}$ can be loosely seen as a generalized tadpole diagram, which evaluates to:
\begin{align}
    \label{eq:gentadpole}
    I_{\nu}^{(n-1)} &= \int\left(\prod_{j=1}^{l} \frac{d^{d} k_{j}}{i \pi^{\frac{d}{2}}}\right) D_{1\ldots n}^{-\nu} = \frac{\Gamma(\nu-l d / 2)}{\Gamma\left(\nu \right)}  \frac{\tilde{\mathcal{U}}^{\nu-(l+1) d / 2}}{\tilde{\mathcal{F}}^{\nu-l d / 2}}\,,
\end{align}
where $\tilde{\mathcal{U}}$ and $\tilde{\mathcal{F}}$ are obtained from $\mathcal{U}$ and $\mathcal{F}$ by letting:
\begin{align}
    \label{eq:feynparrescalings}
    x_1 &\rightarrow x_1^{\prime} = \prod_{i=1}^{n-1} x_i\,,\nonumber\\
    x_j &\rightarrow x_j^{\prime} = (1-x_{j-1}) \prod_{i=j}^{n-1}x_i\quad \text{ for } j = 2,\ldots,n-1\,, \nonumber \\ 
    x_n &\rightarrow x_n^{\prime} = (1-x_{n-1})\,.
\end{align}
In short, this follows from observing that $D_{1\ldots n} = x_1^{\prime} D_1 + \ldots + x_n^{\prime} D_n$. More generally, if we choose to repeatedly combine propagators in some different way, we can look at the generalized tadpole propagator to understand how the Feynman parameters are rescaled from the standard Feynman parametrization. 

Using \eqref{eq:feyntrickgeneral}, we may write
\begin{align}
    \label{eq:fundamentalrecursion}
    I_{\nu_{1}, \ldots, \nu_{n-(\kappa-1)}}^{(\kappa-1)} = \frac{\Gamma(\nu_1+\nu_2)}{\Gamma(\nu_1)\Gamma(\nu_2)}\int_0^1 dx_\kappa\, x_\kappa^{\nu_1-1} (1-x_\kappa)^{\nu_2-1} I_{\nu_{1}+\nu_{2}, \nu_3, \ldots, \nu_{n-\kappa}}^{(\kappa)}\,,
\end{align}
where $\nu_1$ and $\nu_2$ are assumed to be positive. If we iterate the recursion, we obtain
\begin{align}
    I_{\nu_{1}, \ldots, \nu_{n}} &= \frac{\Gamma(\nu)}{\Gamma\left(\nu_{1}\right) \ldots \Gamma\left(\nu_{n}\right)} \left(\prod_{j=1}^{n-1}\int_{0}^{1} d x_{j}\, x_{j}^{\mu_j-1}\left(1-x_{j}\right)^{\nu_{j+1}-1}\right) I_{\nu}^{(n-1)}\,, \nonumber \\
    &= \frac{\Gamma(\nu-l d / 2)}{\Gamma\left(\nu_{1}\right) \ldots \Gamma\left(\nu_{n}\right)}\left(\prod_{j=1}^{n-1} \int_{0}^{1} d x_{j}\, x_{j}^{\mu_{j}-1}\left(1-x_{j}\right)^{\nu_{j+1}-1}\right) \frac{\tilde{\mathcal{U}}^{\nu-(l+1) d / 2}}{\tilde{\mathcal{F}}^{\nu-l d / 2}}\,,
\end{align}
where $\mu_j = \nu_1 + \ldots + \nu_j$, and $\nu = \mu_n$. This result could have also been obtained directly from \eqref{eq:feynmanparametrization} by performing the change of variables in \eqref{eq:feynparrescalings} (including the Jacobian determinant). This makes it clear that \eqref{eq:fundamentalrecursion} implements a recursion identity that leads to (an alternate form of) the Feynman parametrization. Note that we have the following special cases when $\nu_1$ and/or $\nu_2$ is zero
\begin{align}
    \label{eq:recursionspecialcases}
    I_{0,0,\nu_3, \ldots, \nu_{n-(\kappa-1)}}^{(\kappa-1)} &= I_{0,\nu_3, \ldots, \nu_{n-\kappa}}^{(\kappa)}\,, \nonumber \\
    I_{\nu_1,0,\nu_3, \ldots, \nu_{n-(\kappa-1)}}^{(\kappa-1)} &= \lim_{x_{\kappa}\rightarrow 1} I_{\nu_1,\nu_3, \ldots, \nu_{n-\kappa}}^{(\kappa)}\,, \nonumber\\
    I_{0,\nu_2,\nu_3, \ldots, \nu_{n-(\kappa-1)}}^{(\kappa-1)} &= \lim_{x_{\kappa}\rightarrow 0} I_{\nu_2,\nu_3, \ldots ,\nu_{n-\kappa}}^{(\kappa)}\,.
\end{align}
We elaborate on these limits at the end of section \ref{sec:ftapproachregularization}.

\section{Direct integration through differential equations}
\label{sec:directintegrationthroughdiffeqns}
\subsection{Main method}
Now that we have obtained the recursion identities in \eqref{eq:fundamentalrecursion} and \eqref{eq:recursionspecialcases}, we will discuss how each step in the recursion can be performed by solving a system of differential equations. First, we add a set of numerator factors to each family $I_{\vec{\nu}}^{(\kappa)}$, in order to have a complete linearly independent set of propagators for the IBP reduction. Let us denote the set of master integrals by $\vec{I}^{(\kappa)}$. We may then obtain a closed-form linear system of differential equations of the form \cite{Kotikov:1990kg, Remiddi:1997ny}
\begin{align}
    \label{eq:differentialeqIk}
    \partial_{x_{\kappa}} \vec{I}^{(\kappa)} = M_{x_{\kappa}} \vec{I}^{(\kappa)}\,.
\end{align}
Note that the master integrals $\vec{I}^{(\kappa)}$ and the corresponding differential matrix $M_{x_{\kappa}}$, depend on Feynman parameters $x_1,\ldots,x_{\kappa}$, and on the external scales and internal masses of the original integral family. Let us assume we have boundary conditions for the master integrals $\vec{I}^{(\kappa)}$ at a given point in phase-space, and for some numeric values of the Feynman parameters $x_1, \ldots, x_{\kappa}$. In the remainder of the paper, we will typically choose $x_j = 11/23$ for all $j$. The precise choice is not important, as long as the external scales, masses, and $x_j$ do not lie on a singularity of the differential equations. Note that boundary conditions in step $\kappa = n-1$ can be obtained by evaluating \eqref{eq:gentadpole} in a chosen numerical point for the external scales, masses, and Feynman parameters.

We may use \texttt{DiffExp} \cite{Hidding:2020ytt} to obtain a solution for the master integrals $\vec{I}^{(\kappa)}$ in the interval $0<x_{\kappa}<1$, while keeping all other variables at a fixed numerical value. The solution is obtained as a piece-wise collection of generalized series expansions up to a given order, on a covering set of line segments. We seek to use this representation to perform the integrals in \eqref{eq:fundamentalrecursion}. In principle, this is straightforward since we may integrate the series expansions term-by-term and segment by segment in an efficient manner. We may also use the series expansion representation to evaluate the limits in \eqref{eq:recursionspecialcases}. However, there are some obstacles which we discuss next.

The first obstacle is that the integrals which appear on the right-hand side of \eqref{eq:fundamentalrecursion} and  \eqref{eq:recursionspecialcases} may not necessarily correspond to the choice of master integrals $\vec{I}^{(\kappa)}$. Therefore, we should use IBP relations to write all integrals in terms of the given master integrals. Since we have representations for the masters in terms of collections of generalized series expansions, we have to convert the IBP coefficients to the same representation as well, which can be done in an automatic way. 

A second obstacle comes from the (dimensional) regularization of the integrals. In particular, integrals of the form \eqref{eq:fundamentalrecursion} do not converge in general when $\varepsilon$ is close to zero. Instead, it may be necessary to obtain the result by analytically continuing from larger values of $\varepsilon$. Practically, this can be achieved through a formula for resolving the poles \eqref{eq:secdecformula}. Similarly, the limits in \eqref{eq:recursionspecialcases} have to be taken in a way that is consistent with dimensional regularization. We explain in more detail how to deal with these issues in section \ref{sec:ftapproachregularization}. 

The last difficulty is the need for $i\delta$-prescriptions to perform the transport past (threshold) singularities in the bulk of the interval $0<x_{\kappa}<1$. If we choose our kinematics in the Euclidean region, there are no such singularities by definition. If we choose kinematics in physical regions, there are singularities and it is necessary to make choices which are in correspondence with the Feynman prescription. We will discuss this point further in section \ref{sec:threshold}.

Taking care of the previously mentioned subtleties, we may use \eqref{eq:fundamentalrecursion} and \eqref{eq:recursionspecialcases} to obtain new boundary conditions for the master integrals $\vec{I}^{(\kappa-1)}$. By iterating this procedure, we eventually obtain numerical results for the master integrals of the original integral family $\vec{I} = \vec{I}^{(0)}$.

\subsection{Regularization}
\label{sec:ftapproachregularization}
Let us consider the regularization of \eqref{eq:fundamentalrecursion}. In general, Feynman parameter integrals may not converge when $\varepsilon$ is close to zero, due to the presence of non-integrable singularities at the integration boundaries $x_{\kappa} = 0$ and $x_{\kappa}=1$. There may also be singularities in the bulk of the integration region $0<x_{\kappa}<1$, which can be integrated over by deforming the contour using suitable $i\delta$-prescriptions. We will discuss this case in section \ref{sec:threshold}. 

Singularities at the boundary can be dealt with using sector decomposition. What makes our computation different from normal is that our integrand is represented by a collection of series expansions along a partitioning of the interval $x_{\kappa}\in[0,1]$. Let us first distinguish two ways to deal with the dimensional regulator $\varepsilon$ in the series expansion approach. The first option is to work order-by-order in $\varepsilon$. \texttt{DiffExp} was originally written with this approach in mind. However, we found it difficult to apply regularization formulas while the series expansions are in this representation. Instead, we may find a (quasi-)finite basis of master integrals for the differential equations \cite{vonManteuffel:2014qoa}, which can be done using for example the functionality in Reduze 2 \cite{vonManteuffel:2012np}. Non-integrable singularities at the boundary disappear when choosing (quasi-)finite master integrals for each family $I_{\vec{\nu}}^{(\kappa)}$. A downside to this approach is that finite basis integrals may contain high degree numerators and dots, or may be shifted in dimension, and therefore may require more complicated IBP reductions.

The second option for the transport with \texttt{DiffExp} is to use a fixed numeric value for $\varepsilon$. By repeating the computation multiple times for different values of $\varepsilon$, it is possible to reconstruct the dependence on $\varepsilon$ up to a given order \cite{Liu:2022chg}. Although \texttt{DiffExp} has not been designed with this approach in mind, the package also works on a differential matrix with a numeric value plugged in for $\varepsilon$.\footnote{However, because \texttt{DiffExp} uses Mathematica's \texttt{SeriesData} format, it is not able to work with very small values of $\varepsilon$. In particular, the \texttt{SeriesData} representation stores many redundant zero coefficients when a power series is multiplied by an overall fractional power $x^{b\varepsilon}$. For the examples in section \ref{sec:example}, we chose the values in \eqref{eq:epssamplesexplicit}.} The $\varepsilon$-sampling approach is furthermore suitable for analytic regularization of the poles \cite{Panzer:2014gra}, which we discuss next. Let us consider an integral of the form
\begin{align}
    \label{eq:decompositionpiece}
    \int_0^c dx\, x^{a+b\varepsilon} g(x,\varepsilon)\,.
\end{align}
Here we assume that $a$ and $b$ are rational numbers, that $b \neq 0$, and that the upper bound is chosen such that there are no non-integrable singularities in the interval $0<x\leq c$, except possibly at $x = 0$. Furthermore, we assume that $g(x,\varepsilon)$ is a Taylor series in $x$ with a non-zero finite part, and whose coefficients may depend on $\varepsilon$. When $a\leq -1$ and $\varepsilon\rightarrow 0$, there is a non-integrable singularity at $x = 0$. We may resolve it by (possibly repeatedly) applying the following integral identity:
\begin{align}
    \label{eq:secdecformula}
    \int_0^c dx\, x^{a+b\varepsilon} g(x) = \int_0^c dx\, \frac{x^{a+b\varepsilon+1}}{(1+a+b\varepsilon)}\left(\frac{(2+a+b\varepsilon)}{c} g(x)-\left(1-\frac{x}{c}\right)g'(x)\right)\,,
\end{align}
which is similar to the formula for extracting poles in the sector decomposition approach (see e.g. \cite{Heinrich:2008si}). It is clear that this formula increases the overall scaling exponent in the limit $x\rightarrow 0$, such that repeated application of the formula resolves non-integrable singularities.

We discuss next how to apply this formula to integrate results from \texttt{DiffExp}. The solutions from \texttt{DiffExp} are given as a collection of line segments. In particular, there is a segment around $x = 0$, and a segment around $x' = 1 - x = 0$. For illustrative purposes, we will consider the segment around $x=0$, but the discussion is the same for the segment at the upper integration boundary with local line parameter $x'$. We decompose the expansions in the following way:
\begin{align}
    \label{eq:seriesdecomposition}
    g(x) = g_0(x,\epsilon) + x^{a_1+b_1\varepsilon} g_1(x,\varepsilon) + \ldots + x^{a_k+b_k\varepsilon} g_k(x,\varepsilon)\,,
\end{align}
where $g_0(x,\epsilon)$ is a Taylor series in $x$, and where each contribution $x^{a_i+b_i\varepsilon}g_i(x,\varepsilon)$ with $i=1,\ldots,k,$ leads to an integration of the form of \eqref{eq:decompositionpiece}. We may assume that a numerical value for $\varepsilon$ is already plugged into each $g_i(x,\varepsilon)$. The exponents $a_i$ and $b_i$ may be read off from the indicial equations of the differential equations. 

The fact that a decomposition of the form of \eqref{eq:seriesdecomposition} is possible, and that there are no contributions with $a_i \leq -1$ and $b_i = 0$, was established empirically by computing a number of examples. Furthermore, we might expect logarithms in the expansions, but we did not observe these in our examples. (However, logarithms of the type $\log(x)$ may be obtained by expanding in $\epsilon$.) Lastly, we found in our examples that the exponents $a_i$ and $b_i$ were always integers.

After performing the decomposition, we may apply \eqref{eq:secdecformula} at integrand level to each factor in \eqref{eq:seriesdecomposition} until we reach a form where every $a_i \geq 0$. Afterwards, we can perform the integration as usual. Lastly, we remark on how to perform the limits in \eqref{eq:recursionspecialcases} for integrals which do not depend on one of the combined propagators. To evaluate the limits, we take the segment centered at $x = 0$ or at $x'=1-x=0$, and we filter out the finite coefficient of the Taylor series $g_0(x,\epsilon)$ in \eqref{eq:seriesdecomposition}. This means we put any contributions of the form $x^{a_i + b_i \varepsilon} g_i(x,\varepsilon)$ with $b_i \neq 0$ to zero (even when $a_i < 0$).

\subsection{Threshold singularities}
\label{sec:threshold}
When crossing threshold singularities (with \texttt{DiffExp}) it is necessary to provide $i\delta$-prescriptions (contour deformations) which are in agreement with the Feynman prescription. Since our method involves solving differential equations of Feynman parameters, instead of external scales and masses, it is not always clear how to proceed with a suitable choice of $i\delta$. One option is to first compute a point in the Euclidean region (when a Euclidean region exists for the given integral family.) One may then obtain results in the physical region by transporting from the Euclidean region using the system of differential equations of the (original) integral family.

Alternatively, we may seek to compute results directly in the physical region with our method. Let us first recall that if a graph representation exists for an integral family, we may identify possible thresholds by drawing unitarity cuts. In practice, we can use a simple automatic strategy which was also employed in \cite{Dubovyk:2022frj}. We take the set of external legs $p_j$ and compute all possible subsets (including the empty set.) Next, we do the same for the set of all internal masses $m_j$. Lastly, we generate a list of all $i\delta$-prescriptions of the form
\begin{align}
    \label{eq:standardthresholdidelta}
    s - M^2 + i\delta\,,
\end{align}
where $s$ is a sum of momenta, and $M$ is a sum of internal masses. We then provide \texttt{DiffExp} with the resulting set of $i\delta$-prescriptions. The list is overcomplete (and includes pseudo-thresholds), but any redundant prescriptions will be of no effect. For topologies which contain combined propagators, a graphical representation often still exist and we can employ the above strategy on the (generalized) external momenta and masses (cf. \eqref{eq:completingthesquare}). In this case, $s$ and $M$ will depend on the unintegrated Feynman parameter(s).

If we choose to combine only propagators which have the same loop momenta in the first steps of the iteration, we have a diagrammatic representation for the integral families in these steps. For example, for the two-loop examples in section \ref{sec:example} we eventually obtain a (generalized) sunrise topology (see Figure \ref{fig:doublePentagonx5}). In the final steps of the iteration, we have to combine propagators with different loop momenta, and here we do not currently have a general understanding of which $i\delta$-prescriptions to give to \texttt{DiffExp}. We note that we do not always cross physical thresholds for every integration. For example, in section \ref{sec:exampledoublepentagon} we compute a non-planar double pentagon family in a kinematic point that lies outside the Euclidean region. However, we found that we do not cross any physical thresholds during the transports of $x_4, x_5, x_6$ and $x_7$. We elaborate on this point in section \ref{sec:exampledoublepentagon}.

More generally, we may need to cross physical thresholds of integral families that do not have a diagrammatic depiction. On the one hand, we could choose to compute boundary conditions for the last diagrammatic topology in the iteration using another method such as auxiliary mass flow, or by deriving boundary conditions manually using expansion by regions. On the other hand, we believe that a suitable contour prescription exist in principle for each Feynman parameter. For example, we managed to obtain consistent results in the physical region of the sunrise integral family with our method, after trying a few different choices of $i\delta$-prescriptions. To get consistent results, we also included a small $-i\delta$ in the second Symanzik polynomial appearing in the boundary conditions of the generalized tadpole topology (cf. \eqref{eq:symanzikdef}, \eqref{eq:gentadpole}).

We hope that a more rigorous analysis in the future will shed further light on the general choice of $i\delta$-prescriptions for the Feynman parameters, especially when computing integral families that do not have a diagrammatic representation.

\section{Pedagogical examples}
\label{sec:example}
\begin{figure}[h]
\begin{center}
\includegraphics[scale=0.75]{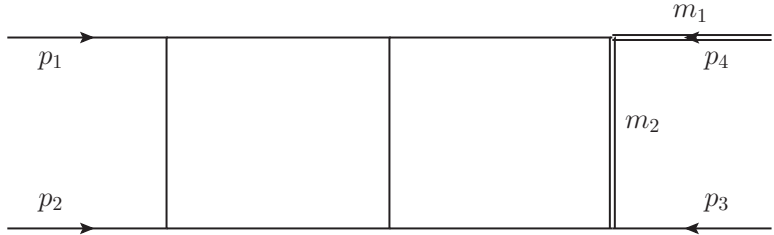}
\caption{ 
A double box topology named \text{topo7} which in this paper is dressed with one massive external leg and one massive propagator.
}
\label{fig:doublebox}
\end{center}
\end{figure}

In this section, we define two integral families which we will use for our examples. The computation of the second integral family will be discussed in detail in section \ref{sec:exampledoublepentagon}. For the examples, we have opted to flip the sign of the momentum and mass in the propagators compared to section \ref{sec:preliminaries}. The first example family is defined by
\begin{align}
    I^{\text{topo7}}_{\nu_1, \nu_2, \nu_3, \nu_4, \nu_5, \nu_6, \nu_7, \nu_8, \nu_9} = \int \frac{d^dk_{1}}{i\pi^{\frac{d}{2}}}\frac{d^{d}k_{2}}{i\pi^{\frac{d}{2}}}\, \frac{ D_8^{-\nu_8}D_9^{-\nu_9}}{D_1^{\nu_1}D_2^{\nu_2}D_3^{\nu_3}D_4^{\nu_4}D_5^{\nu_5}D_6^{\nu_6}D_7^{\nu_7}}\,,
\end{align}
where $d=4-2\varepsilon$, and where the propagators are given by
\begin{align}
D_{1} &= (k_1+p_3)^2 - m_2^2\,, & D_{4} &= k_2^2\,, & D_{7} &= (k_1-k_2)^2\,, \nonumber\\
D_{2} &= k_1^2\,, & D_{5}\, &= (k_2-p_2)^2\,, & D_{8} &= (k_1-p_2)^2\,, \nonumber\\
D_{3} &= (k_1 - p_1 - p_2)^2\,, & D_{6} &= (k_2 - p_1 - p_2)^2\,, & D_{9} &= (k_2-p_3)^2\,,
\end{align}
where $D_8$ and $D_9$ are assumed to be numerators ($\nu_8, \nu_9 \leq 0$). The top sector is depicted in Figure \ref{fig:doublebox}. The kinematics is given by
\begin{align}
    p_{1}^2 = p_{2}^2 = p_{3}^2 = 0\,,  \quad  \quad  p_{1}\cdot p_{2} = s/2\,, \quad \quad  p_{1}\cdot p_{3} = (t-s+m_{1}^{2})/2\,, \quad \quad p_{2}\cdot p_{3} = -t/2\,,
\end{align}
where the momenta satisfy the conservation identity $p_1+p_2+p_3+p_4=0$.

\begin{figure}[h]
\begin{center}
\includegraphics[scale=0.75]{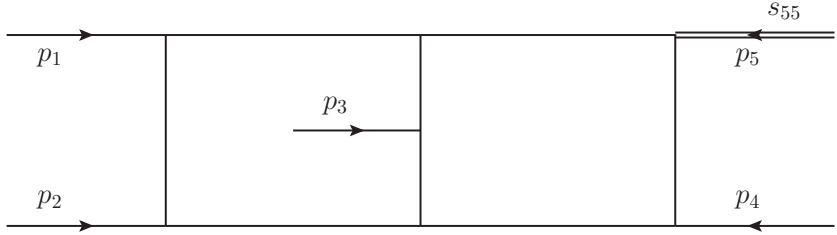}
\caption{ 
A double pentagon topology named \text{5p} which in this paper is dressed with one massive external leg.
}
\label{fig:doublePentagon}
\end{center}
\end{figure}

\noindent Our second example is of higher complexity, and defined by
\begin{align}
     \label{eq:doublepentagon}
     I^{\text{5p}}_{\nu_1, \nu_2, \nu_3, \nu_4, \nu_5, \nu_6, \nu_7, \nu_8, \nu_9, \nu_{10},\nu_{11}} = \int  \frac{d^dk_{1}}{i\pi^{\frac{d}{2}}}\frac{d^{d}k_{2}}{i\pi^{\frac{d}{2}}}\,\frac{D_9^{-\nu_9}D_{10}^{-\nu_{10}}D_{11}^{-\nu_{11}}}{D_1^{\nu_1}D_2^{\nu_2}D_3^{\nu_3}D_4^{\nu_4}D_5^{\nu_5}D_6^{\nu_6}D_7^{\nu_7}D_8^{\nu_8}}\,,
\end{align}
where $d=4-2\varepsilon$, and where the propagators are
\begin{align}
D_{1}&=\left(k_{2}-p_{1}-p_{2}-p_{3}-p_{4}\right)^{2}\,, & D_{5}&=\left(k_{1}-p_{1}\right)^{2}\,, & D_{9}&=\left(k_{2}-p_{1}-p_{2}\right)^{2}\,, \nonumber\\
D_{2}&=\left(k_{2}-p_{1}-p_{2}-p_{3}\right)^{2}\,, & D_{6}&=k_{1}^{2}\,, & D_{10}&=\left(k_{1}-p_{1}-p_{2}-p_{3}-p_{4}\right)^{2} \,,\nonumber\\
D_{3}&=k_{2}^{2}\,, & D_{7}&=\left(k_{1}-k_{2}+p_{3}\right)^{2}\,, & D_{11}&=\left(k_{2}-p_{1}\right)^{2}\,, \nonumber\\
D_{4}&=\left(k_{1}-p_{1}-p_{2}\right)^{2}\,, & D_{8}&=\left(k_{1}-k_{2}\right)^{2}\,.
\end{align}
We assume that $D_9$, $D_{10}$ and $D_{11}$ are numerators ($\nu_9, \nu_{10}, \nu_{11} \leq 0$). The kinematics is given by:
\begin{align}
    &p_1^2=p_2^2=p_3^2=p_4^2=0\,,\quad\quad p_1\cdot p_2 = s_{12}/2\,,\quad\quad p_1\cdot p_3 = s_{13}/2\,,\quad\quad p_1\cdot p_4 = s_{14}/2\,, \nonumber \\ 
    &p_2\cdot p_3 = s_{23}/2\,,\quad\quad p_2\cdot p_4 = -(s_{12} + s_{13} + s_{14} + s_{23} + s_{34} - s_{55})/2\,,\quad\quad p_3\cdot p_4 = s_{34}/2\,,
\end{align}
where the momenta are conserved according to $p_1+p_2+p_3+p_4+p_5=0$. The family is depicted in Figure \ref{fig:doublePentagon}.

\subsection{Non-planar double pentagon family}
\label{sec:exampledoublepentagon}
In the following section we describe the application of our method to the non-planar double pentagon family $I^{\text{5p}}_{\vec{\nu}}$, with a focus on the computational aspects of the calculation. We obtained our results using the $\varepsilon$-sampling strategy, which we borrowed from \cite{Liu:2022chg}. More specifically, we considered a set of small numerical values $\{\varepsilon_i\}$ inserted for the dimensional regulator $\varepsilon$, and we performed the computation every time for a fixed $\varepsilon_i$. At the end, we reconstructed our results as a power series in $\varepsilon$ up to order $\varepsilon^4$. 

We aimed to reach a precision of 40 digits for the finite part of all master integrals of the uncombined integral topology defined in \eqref{eq:doublepentagon}. To do so, we proceeded with some small trial and error. We found it sufficient to choose numerical values 
\begin{align}
    \label{eq:epssamplesexplicit}
    \varepsilon_i=(-1)^i/(50+i) \quad \text{where $i=1,\dots, 40$\,,}
\end{align}
and to perform the evaluation of all integrals at a precision of 60 digits. Our samples $\varepsilon_i$ are relatively large in magnitude, because \texttt{DiffExp} can currently not handle very small samples. Note that when reconstructing the dependence on $\varepsilon$ up to a given order, there is a precision loss involved. We reconstructed our results a few times by leaving out a sample from the full set. By comparing the different results, we found a variation that was within the desired precision of 40 digits for the finite part. A more systematic analysis for how to choose the precision of the evaluations, the values for $\varepsilon_i$, and the number of samples, is described in \cite{Liu:2022chg}.

\begin{figure}[h]
\begin{center}
\includegraphics[scale=0.75]{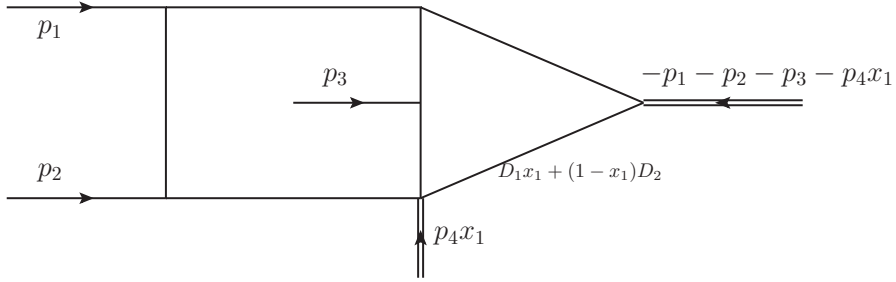}
\caption{ 
Step $\kappa = 1$: Combining propagators for the first time, reducing the complexity of the initial topology 5p.
}
\label{fig:doublePentagonx1}
\end{center}
\end{figure}

\begin{figure}[h]
\begin{center}
\includegraphics[scale=0.75]{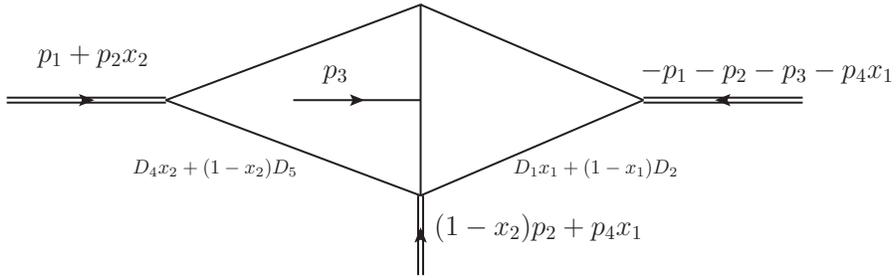}
\caption{Step $\kappa = 2$: Combining propagators for the second time, reducing the complexity further.
}
\label{fig:doublePentagonx2}
\end{center}
\end{figure}

\begin{figure}[h]
\begin{center}
\includegraphics[scale=0.75]{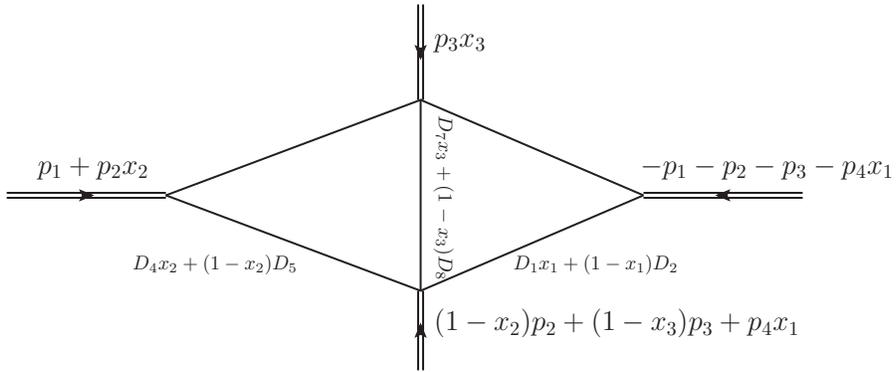}
\caption{Step $\kappa = 3$: Combining propagators for the third time, reducing the complexity further.
}
\label{fig:doublePentagonx3}
\end{center}
\end{figure}

\begin{figure}[h]
\begin{center}
\includegraphics[scale=0.75]{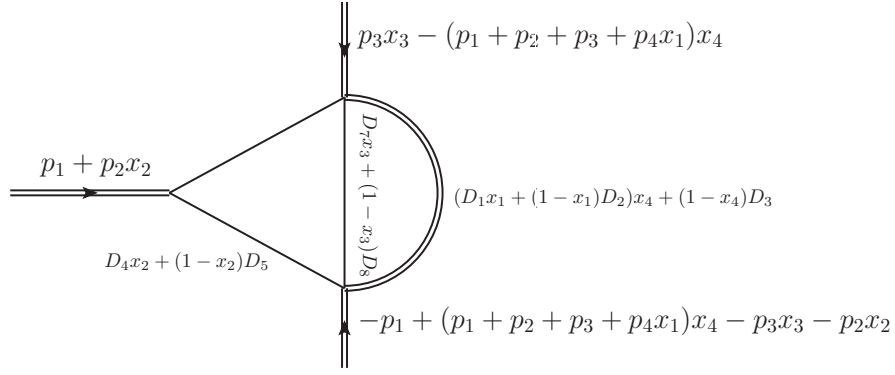}
\caption{Step $\kappa = 4$: Combining propagators for the fourth time, reducing the complexity further.
}
\label{fig:doublePentagonx4}
\end{center}
\end{figure}

\begin{figure}[h]
\begin{center}
\includegraphics[scale=0.8]{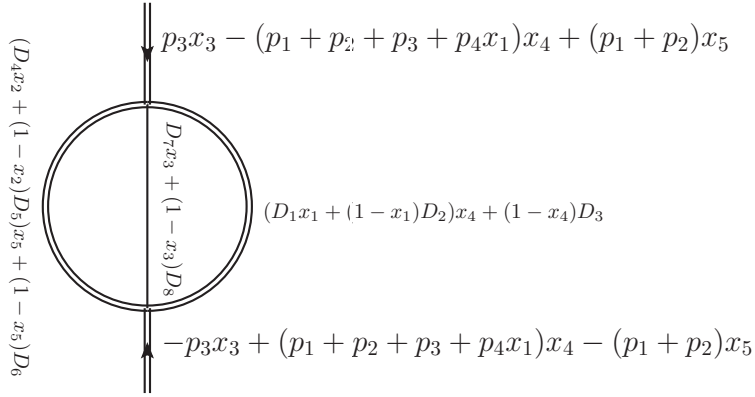}
\caption{Step $\kappa = 5$: Combining propagators for the fifth time, reducing the complexity further, and giving a sunrise topology. Further iterations do not have a diagrammatic representation.
}
\label{fig:doublePentagonx5}
\end{center}
\end{figure}

{\renewcommand{\arraystretch}{1.5}
\begin{table}[h]
\centering
\caption{We iterate Feynman's trick seven times to reduce the non-planar double pentagon topology to a trivial form. Each step $\kappa$ combines two propagators in the input column. The new combined propagator is given in the output column. The number of master integrals of the integral family with the combined propagator(s) is given in the last column. A large reduction of master integrals is seen for each iteration.}
\begin{tabular}[c]{|c|c|c|c|}
\hline
$\kappa$ & input & output & Number of master integrals  \\\hline
0 & - & uncombined & 142\\
1 & $\{D_1,D_2\}$& $D_{12}=D_1x_1+(1-x_1)D_2$ & 69 \\
2 & $\{D_4,D_5\}$ & $D_{45}=D_4x_2+(1-x_2)D_5$ & 32 \\
3 & $\{D_7,D_8\}$ & $D_{78}=D_7x_3+(1-x_3)D_8$ & 16\\
4 & $\{D_{12}, D_3\}$ & $D_{123}=D_{12} x_4 +(1-x_4)D_3$ & 8 \\
5 & $\{D_{45}, D_6\}$ & $D_{456}=D_{45} x_5 +(1-x_5)D_6$ & 4 \\
6 & $\{D_{123},D_{456}\}$ & $D_{123456}=D_{123}x_6+(1-x_6)D_{456}$ & 2 \\
7 & $\{D_{123456},D_{78}\}$ & $D_{12345678}=D_{123456}x_7+(1-x_7)D_{78}$ & 1 \\\hline
\end{tabular}
\label{tab:iteration}
\end{table}}
Our computation required seven iterations, where we repeatedly combine two propagators. We provide the order in which the propagators are combined in Table \ref{tab:iteration}. The ancillary file \texttt{5-point-FT-iteration.wl} contains the definitions of the topologies with combined propagators. There exists a diagrammatic representation for the integral families in steps $\kappa=0,\dots,5$, which are depicted in figures \ref{fig:doublePentagon}, \ref{fig:doublePentagonx1}, \ref{fig:doublePentagonx2}, \ref{fig:doublePentagonx3}, \ref{fig:doublePentagonx4}, \ref{fig:doublePentagonx5}. In step $\kappa = 5$, we have a generalized sunrise family, which is computed with two further iterations where the topologies do not have a diagrammatic representation anymore. The integral family $I_{\nu}^{\text{5p}, (\kappa=7)}$ in step $\kappa = 7$ can be loosely viewed as a generalized tadpole with a single master integral, which evaluates to a combination of (rescaled) Symanzik polynomials and gamma functions (cf. \eqref{eq:gentadpole}). We obtain the first boundary conditions by setting all external scales and Feynman parameters to numerical values. In particular, we consider the point $s_{14}= 3$, $s_{13} = -11/17$, $s_{23} = -13/17$, $s_{12} = -7/17$, $s_{34} = -7/13$, $s_{55} = -1$, and we set all Feynman parameters to the value $x_j = 11/23$. We then use \texttt{DiffExp} to transport in $x_7$, and to obtain a generalized series representation in the interval $0<x_7<1$. By integrating the series representation in accordance with Feynman's trick \eqref{eq:feyntrick}, we obtain boundary conditions for the integral family in step $\kappa=6$. We repeat the procedure of transporting and integrating, until we reach the last Feynman parameter integration in step $\kappa=1$. 

After performing the last integration, we obtain numerical results for all master integrals of the original family ($\kappa = 0$) defined in \eqref{eq:doublepentagon}. These results may also be used as boundary conditions for the differential equations of the original integral family, which allows one to reach other points in phase space without going through the iteration again. We remark that we did not cross any physical thresholds for the first transports in the Feynman parameters $x_7,\ldots,x_4$. This can be motivated by the fact that the second Symanzik polynomial, obtained from the generalized tadpole in step $\kappa = 7$, is positive in our kinematic point with $x_i = 11/23$ for $i = 1,2,3,$ and with $0<x_j<1$ for $j=4,\ldots,7$. For the transports of $x_3$ and $x_1$ we found that we have to cross physical thresholds, while for the transport in $x_2$ we did not observe any physical thresholds. We chose $i\delta$-prescriptions for $x_3$ and $x_1$ in the manner described in section \ref{sec:threshold}. The result for one of the 142 most complicated master integrals is given below at the numerical point $s_{14}= 3$, $s_{13} = -11/17$, $s_{23} = -13/17$, $s_{12} = -7/17$, $s_{34} = -7/13$, $s_{55} = -1$:
\begin{align}
    I^{\text{5p}}_{13111111000}& = \frac{1}{\varepsilon^{-4}}\big(-80991.44634941832815855134956686330134244459\big)  \nonumber\\
 +&\frac{1}{\varepsilon^{-3}}\big(-1176854.140501650857516200908950071824160111- \nonumber\\ &\quad\quad 303701.8453350029342400125918254935316349429i\big)  \nonumber\\
 +&\frac{1}{\varepsilon^{-2}}\big(-13432835.8477692962185637394931604891797674- \nonumber\\ &\quad\quad 4251651.64965980166114774272201533676580580i\big)  \nonumber\\
 +&\frac{1}{\varepsilon^{-1}}\big(-111346171.63704503288070435527859004232921- \nonumber\\ &\quad\quad 32927342.395688330300021665788556801968176i\big)  \nonumber\\
 +&\big(-763045644.5561305442093867867513427731742- \nonumber\\ &\quad\quad 183231121.4048774146788661490531205282119i\big)  \nonumber\\
 +&\varepsilon\big(-4428755434.16119754697555927652734791719- \nonumber\\ &\quad\quad 816059490.912195429388068459166197648719i\big)  \nonumber\\
 +&\varepsilon^2\big(-23085640630.259889520777994526537639199- \nonumber\\ &\quad\quad 3082908606.7551294811504215473642629605i\big)  \nonumber\\
 +&\varepsilon^3\big(-110164352209.7092412652451256610943938- \nonumber\\ &\quad\quad 10252510409.42185691550687766152353640i\big)  \nonumber\\
 +&\varepsilon^4\big(-497649560130.015209279192098631531920- \nonumber\\ &\quad\quad 30796992268.3516086870566559550754104i\big).
\end{align}
We have numerically cross-checked the master integrals with \texttt{AMFlow}.

\subsection{Computational complexity}
The main advantage of our method is that the integral families with combined propagators typically have significantly fewer master integrals for each step in the iteration. We also remark that there is a significant overlap between the master integrals of different steps in the iteration. This feature could be utilized in a future specialized integration code. We motivate with four different examples in Table \ref{tab:compare} that the computational complexity of our method is in many cases reduced compared to auxiliary mass flow \cite{Liu:2022chg}. In particular, if we assume that the same algorithms are used to perform the IBP reductions and the transportation of boundary conditions from the differential equations, we expect that the computational complexity is mostly determined by the maximal number of master integrals appearing in the computation of the Feynman integrals. Our example family \text{5p} serves as an ideal candidate to showcase a significant reduction in the computational complexity compared to the auxiliary mass flow method. We found in this case that the CPU hours for performing the IBP-reductions are reduced by a factor of 66 between our method and auxiliary mass flow.
{\renewcommand{\arraystretch}{1.5}
\begin{table}[h]
\centering
\caption{The maximal number of master integrals appearing in the computation of a few integral families.}
\begin{tabular}[c]{|c|c|c|c|}
\hline
Integral family & No deformation & Combined propagators & \texttt{AMFlow}  \\\hline
topo7 & 31 & 19 & 31 \\
topo7 with  $m_1=0$, $m_2=0$  & 8 & 12 & 21 \\
5p & 142 & 69 & 191 \\
5p with $s_{55}=0$ & 108 & 69 & 174 \\\hline
\end{tabular}
\label{tab:compare}
\end{table}}

There are a few possible caveats to the above analysis. Firstly, in our current implementation it takes relatively long to compute the series expansions with \texttt{DiffExp} in the intervals $0<x_j<1$. Furthermore, the integration and regularization of the series expansions (cf. \eqref{eq:secdecformula} and \eqref{eq:fundamentalrecursion}) takes some time in our current implementation. For example, the final transport with \texttt{DiffExp} and the subsequent integration took 340 CPU hours in total for the \text{5p} integral family. The whole bottom-up iteration for the \text{5p} integral family took 700 CPU hours (which includes the computation of all 40 numerical samples in the dimensional regulator.) We expect that incorporating ideas from the differential solver used in \texttt{AMFlow} will improve the performance of \texttt{DiffExp} significantly.

\section{Conclusion}
\label{sec:conclusion}
In this work we demonstrated a novel method for computing Feynman integrals. The method is similar to the direct integration method, in which the Feynman parametrization is integrated one Feynman parameter at a time. However, in our approach each integration is done numerically by solving a system of differential equations of a simplified Feynman integral family. The solutions are obtained in terms of generalized series expansions, which are subsequently integrated term-by-term. We furthermore showed that non-integrable singularities at the integration boundaries can be dealt with using a regularization formula. This leads to a computationally efficient and precise method for performing Feynman parameter integrals numerically. Lastly, we have made a first exploration of how to obtain results outside the Euclidean region with our method. (Although a full understanding of the necessary contour deformations of the Feynman parameters is left for the future.)

The approach discussed in this paper gives a fully algorithmic way to numerically compute master integrals belonging to a generic Feynman integral family. Furthermore, we expect that our method is computationally more efficient than auxiliary mass flow, as it typically requires the computation of integral families with fewer master integrals (see Table \ref{tab:compare}). This significantly reduces the complexity of the IBP reductions which are required. As a proof of concept, we have discussed in detail the computation of a non-planar double pentagon integral family using our method. We leave the publication of a general public code that implements our method to a future publication.

\section*{Acknowledgements}
JU and MH thank Xiao Liu for fruitful discussions during the 2 weeks workshop \textit{Precision calculations for future e+e– colliders: targets and tools} at CERN. JU and MH thank Oliver Schlotterer and Stefan Weinzierl for short comments on the manuscript. JU thanks Janusz Gluza and his group for providing computing resources with the support in part by the Polish National Science Center (NCN) under grant 2017/25/B/ST2/01987. MH is supported by the European Research Council under ERC-STG-804286 UNISCAMP.
\clearpage
\section*{Appendix A: Feynman trick iterations of the double pentagon family}
\label{sec:apendix}
\clearpage
\bibliographystyle{utphys}
\providecommand{\href}[2]{#2}\begingroup\raggedright\endgroup

\end{document}